\documentclass[lettersize,journal]{IEEEtran}
\usepackage{amsmath,amsfonts}
\usepackage{algorithmic}
\usepackage{algorithm}
\usepackage{array}
\usepackage[caption=false,font=normalsize,labelfont=sf,textfont=sf]{subfig}
\usepackage{textcomp}
\usepackage{stfloats}
\usepackage{url}
\usepackage{verbatim}
\usepackage{graphicx}
\usepackage{cite}
\usepackage{siunitx}

\usepackage{subfiles} 
\usepackage{commath}
\usepackage{multirow}
\usepackage{nicematrix}
\usepackage{color}
\usepackage{tablefootnote}
\usepackage{bbding}
\usepackage{amsmath}

\usepackage{threeparttable}
\usepackage{tablefootnote}

\usepackage{fancyhdr} 
\begin{document}

\newcommand\textline[4][t]{%
  \par\smallskip\noindent\parbox[#1]{.4\textwidth}{\raggedright#2}%
  \parbox[#1]{.2\textwidth}{\centering#3}%
  \parbox[#1]{.4\textwidth}{\raggedleft\texttt{#4}}\par\smallskip%
}

\makeatletter

\def\ps@IEEEtitlepagestyle{%
  \def\@oddfoot{\mycopyrightnotice}%
  \def\@evenfoot{}%
}
\def\mycopyrightnotice{%
 \textline[t]{979-8-3315-1886-8/25/\$31.00 \copyright 2025 IEEE}{\thepage}{}
  \gdef\mycopyrightnotice{}
}

\title{
Physics-Informed Neural Network for Parameter Identification: a Buck Converter Case Study
}

\author{Shuyu Ou,~\IEEEmembership{Student Member,~IEEE,}
Subham Sahoo,~\IEEEmembership{Senior Member,~IEEE,}
Ariya Sangwongwanich,~\IEEEmembership{Senior Member,~IEEE,}
Frede Blaabjerg,~\IEEEmembership{Fellow,~IEEE,}
Mahyar Hassanifar,~\IEEEmembership{Student Member,~IEEE,}\\
Martin Votava,~\IEEEmembership{Member,~IEEE,}
Marius Langwasser,~\IEEEmembership{Member,~IEEE,} and
Marco Liserre,~\IEEEmembership{Fellow,~IEEE} 
}
\markboth{ 
}%
{Shell \MakeLowercase{\textit{et al.}}: A Sample Article Using IEEEtran.cls for IEEE Journals}


\maketitle
\begin{abstract}

System-level condition monitoring methods estimate the electrical parameters of multiple components in a converter to assess their health status. The estimation accuracy and variation can differ significantly across parameters. For instance, inductance estimations are generally more accurate and stable than inductor resistance in a buck converter. However, these performance differences remain to be analyzed with a more systematic approach otherwise the condition monitoring results can be unreliable. 
Therefore, this paper analyzes the training loss landscape against multiple parameters of a buck converter to provide a systematic explanation of different performances. If the training loss is high and smooth, the estimated circuit parameter typically is accurate and has low variation. 
Furthermore, a novel physics-informed neural network (PINN) is proposed, offering  {\color{black}faster convergence and lower computation requirements compared to an existing PINN method}. The proposed method is validated through simulations, where the loss landscape identifies the {\color{black}unreliable} parameter estimations, and the PINN can estimate the remaining parameters. 



\end{abstract}

\vspace{-15pt}
\section{Introduction}
Condition monitoring methods are used on power electronics converters to identify degradation and prevent failures. A recent trend in research focuses on system-level condition monitoring methods that utilize terminal measurements to estimate key circuit parameters. These methods are non-invasive and allow for monitoring multiple components (e.g., inductances, capacitances, and resistances) with a single method \cite{Sihui_1pInverter_2023, Sihui_3pInverter_2024}. 

However, inconsistent performance occurs in multi-parameter estimation \cite{peng_digital_2021}. For instance, in a buck converter, the inductance $L$ can be accurately estimated; while the inductor resistance $R_\mathrm{L}$ and MOSFET on-state resistance $R_\mathrm{dson}$ are difficult to estimate, because they have lower impacts on the state vector \cite{She2024} and their series connection creates coupling in the state space model \cite{zhao_parameter_2022}. 
Various methods have been proposed to address the problem, such as incorporating an additional neural network~\cite{She2024}, decoupling the parameters based on the switching state \cite{Choksi2023}, or adding physics models \cite{10712655}. 

{\color{black}Nonetheless, the fundamental question remains: which parameters are difficult to estimate? Without such analysis, the risk of deriving unreliable estimations exists, leading to false warnings and unnecessary maintenance actions \cite{black2021condition}}. 


Therefore, this paper analyzes the relationship between the training loss landscape and the performances in multi-parameter estimations. The results are studied from the magnitude and smoothness of the loss curve, 
where a higher magnitude and a smoother curve generally indicate accurate and low-variation estimations that are feasible for condition monitoring. The loss landscape study provides a systematic approach for evaluating the system-level condition monitoring method, for instance, in the buck converter case study, several parameters have inaccurate and high-variation estimations that can be explained using the loss landscape, indicating the need for additional inputs or extra physics models \cite{She2024, Choksi2023, 10712655}. 

Additionally, various optimizers have been applied to estimate circuit parameters~\cite{zhiwei_ANPC_2024, Gabriel2021LS}, 
within which, the physics-informed neural network (PINN) has stabler estimations so it is selected \cite{She2024}. 
The PINN structure is further modified in terms of network size and network outputs, achieving {\color{black}smaller neural network, faster convergence, and reduced computational requirement.} 

In summary, the main contributions are: 1) analyzing the loss landscape to {\color{black}systematically evaluate the system-level condition monitoring in multi-parameter estimations, with which, unreliable estimations are identified and unnecessary maintenance actions can be avoided}; 
and 2) proposing a new PINN structure with {\color{black}lower computational requirements.}  

The rest of the paper is organized as follows: Sections II and III introduce the buck converter and the proposed method. Section IV analyzes the method performance and training loss landscape through simulation. Finally, Section V concludes the proposed method and results.

\section{System Description}

\begin{figure}[!t]
\centering
\includegraphics[width=2.8 in, page=1]{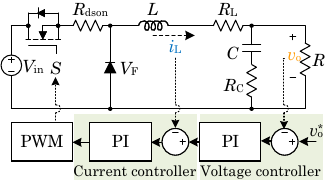}
\caption{{\color{black}A buck converter controlled with voltage and current controllers, where PI is the proportional-integral controller and PWM is the pulse-width modulation.}}
\label{fig_buck}
\end{figure}
The proposed method is demonstrated on a buck converter, as illustrated in Fig. \ref{fig_buck}, comprising a MOSFET, a diode, an inductor $L$, and a capacitor $C$. 
{\color{black}The load resistance is $R$. The on-state resistance of MOSFET is $R_\mathrm{dson}$, and the diode forward voltage drop is $V_\mathrm F$. The equivalent series resistances of inductance and capacitance are $R_\mathrm L$ and $R_\mathrm C$, respectively.}

The converter operates in current control mode,  regulating both the inductor current $i_\mathrm L$ and output voltage $v_\mathrm o$ by adjusting the duty ratio of the switching signal $S$. 


The circuit parameter being estimated are $\theta_0$=$\{L$, $C$, $R$, $R_\mathrm{C}$, $R_\mathrm{L}$, $R_\mathrm{dson}$, $V_\mathrm{F}\}$, 
{\color{black}but only the first four terms $\theta_1$=$\{L$, $C$, $R$, $R_\mathrm{C}\}$ are used for condition monitoring due to their accurate and robust estimations, as demonstrated in the result section.} In contrast, the remaining parameters $\theta_2=$$\{R_\mathrm{L}$, $R_\mathrm{dson}$, $V_\mathrm{F}\}$ have inaccurate and high-variation estimations.

The basic assumptions of this work are:
\begin{itemize}
    \item \textbf{Operating mode}: The converter operates in continuous-conduction mode (CCM). Another mode, discontinuous-conduction mode (DCM), can also be included but requires additional modeling effort \cite{DCM_2025}.
    \item \textbf{Switching signal}: The switching signal is accurately reconstructed using low-sampling rate measurement, such as the method in \cite{zhao_parameter_2022}.   
\end{itemize}

\section{Proposed Method}
The key concept of the proposed method is illustrated in Fig.~\ref{fig_method}, which consists of three functional blocks: a neural network, state vector prediction, and loss calculation.

The neural network estimates the circuit parameters $\hat{\theta}_0$. Then the prediction block uses $\hat{\theta}_0$ to build the state-space model and predicts the state vector within the sampling window. The training loss, computed as the difference between the predicted and measured state vector, is used to update $\hat{\theta}_0$. 
Details of each block are provided in the following sections. 

\subsection{Neural Network}
\begin{figure}[!t]
\centering
\includegraphics[width=3.30 in, page=2]{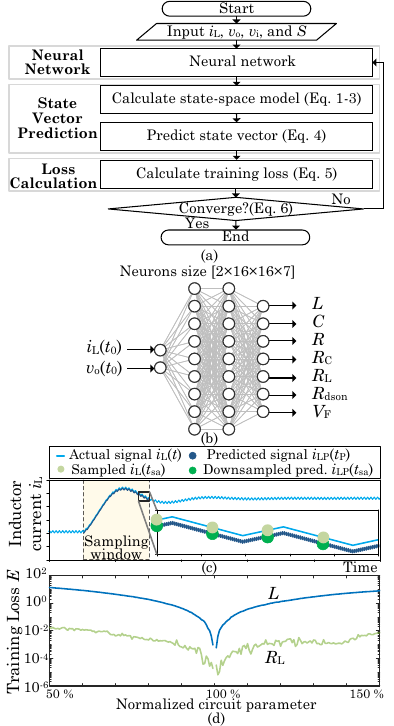}
\vspace{-8 pt}
\caption{ (a) A flowchart of the proposed method. (b) The neural network. (c) Example waveforms of the actual, sampled, predicted, and downsampled predicted state variable $i_\mathrm{L}$. {\color{black}(d) Example of the training loss curves against the inductance $\hat{L}$ and inductor resistance $\hat{R}_\mathrm{L}$.}}
\label{fig_method}
\end{figure}

As illustrated in Fig.~\ref{fig_method}(b), the neural network comprises four layers, with [2$\times$16$\times$16$\times$7] neurons in each layer. {\color{black} The neuron configuration is designed to ensure the minimum number of neurons necessary for stable estimations.} 
The neural network takes inputs of the initial state vector and outputs normalized circuit parameters, $\hat{\theta}_0$~=~$\{L$, $C$, $R$, $R_C$, $R_\mathrm L$, $R_\mathrm{dson}$, $V_\mathrm F\}$. The activation function is $tanh$ for the first three layers, while the last layer uses a $sigmoid$ function to clamp the normalized circuit parameters between zero and one.
\subsection{State Vector Prediction}
The estimated circuit parameters $\hat{\theta}_0$ are used to build a state-space model~\eqref{eq_statespace_on},
\begin{equation}
\label{eq_statespace_on}
\frac{d x}{d t}= Ax+Bu, 
\end{equation}
where $A$, $B$, $x$, and $u$ represent the state matrix, input matrix, state vector, and input vector, respectively. Specifically, the state vector is defined as $x = [i_\mathrm L, v_\mathrm o]^T$ and the input vector is $u = [S, \bar{S}]^T$. When the MOSFET is turned on, $S=1$ and $\bar{S} = 0$, while in the off state, $S=0$ and $\bar{S} = 1$.
\begin{equation}
\label{eq_statespace_A}
A = \begin{bmatrix}
 -\frac{SR_\mathrm{dson}+R_\mathrm L}{L}& -\frac{1}{L}\\
\frac{R L+CRR_\mathrm C(SR_\mathrm{dson}+R_\mathrm L)}{LC(R+R_\mathrm C)}& -\frac{CR R_\mathrm C + L}{LC(R+R_\mathrm C)}\\
\end{bmatrix}
\end{equation}
\begin{equation}
\label{eq_statespace_B}
B =  \begin{bmatrix}
\frac{V_\mathrm{i}}{L}&-\frac{V_\mathrm F}{L}\\ \frac{R_\mathrm CRV_\mathrm {i}}{L(R+R_\mathrm C)} & \frac{-R_\mathrm C RV_\mathrm F}{L(R+R_\mathrm C)}\\
\end{bmatrix}
\end{equation}

The state vector within the sampling period is then predicted using the Forward Euler method via~\eqref{eq_forwardeuler}, where $x_\mathrm{P}$, $k$, $M$, $\Delta t_\mathrm{P}$ represent the predicted state vector, time step index, number of prediction steps, and prediction step size.
\begin{equation}
\label{eq_forwardeuler}
    x_\mathrm{P}(t_\mathrm{Pj})=\left\{ 
    \begin{array}{@{}l@{}l}
    x_\mathrm{P}(t_\mathrm{saj})&,j = 1\\
    x_\mathrm{P}(t_\mathrm{Pj-1}) + \frac{d x_\mathrm{P}(t_\mathrm{Pj-1})}{d t}\Delta t_\mathrm{P}&,j = 2, 3, ..., M
    \end{array} 
    \right.
\end{equation}

\subsection{Training Loss and Neural Network Optimization}
The training loss is defined as the mean squared error (MSE) between the predicted state vector $x_\mathrm P(t_\mathrm P)=[i_\mathrm{LP}(t_ \mathrm P), v_{oP}(t_\mathrm P)]^T$ and the measured state vector $x(t_{sa}) = [i_\mathrm {L}(t_\mathrm{sa}), v_\mathrm{o}(t_\mathrm{sa})]^T$. 

The prediction and sampling time series are defined as $t_\mathrm P$=$\{t_\mathrm{P1}$, $t_\mathrm{P2}$, ..., $t_\mathrm{PM}\}$ and $t_\mathrm{sa}$=$\{t_\mathrm{sa1}$, $t_\mathrm{sa2}$, ..., $t_\mathrm{saN}\}$, respectively, as illustrated in Fig.~\ref{fig_method}(c). The number of prediction steps (dark blue dots) is designed to be significantly greater than the sampling steps (light green dots) to reduce the truncation error \cite{zhao_parameter_2022}. 
To align with sampled data and compute the training loss, the predicted state vector must be downsampled from $M$ to $N$.

\begin{equation}    
\begin{split}
 E(\hat{\theta}_0) = \frac{1}{N} 
   \sum_{j=1}^{N}[ x_\mathrm{P}(t_\mathrm{saj}) - x_\mathrm{sa}(t_\mathrm{saj})]^2
      \label{eq_loss} 
\end{split}
\end{equation}

With the training loss $E$, optimizers update the weights and biases of the neural network. The optimization process utilizes an adaptive moment estimation (Adam) optimizer for initial optimization, followed by a limited-memory Broyden-Fletcher-Goldfarb-Shanno (L-BFGS) optimizer for fine-tuning~\cite{zhao_parameter_2022, taylor2022optimizing}. 

The convergence criteria are defined as reaching a predefined tolerance $\epsilon_\mathrm{\theta_0}$ or the maximum epoch $k$.
\begin{equation}
    {\color{black} \Delta \hat{\theta}_\mathrm{0} < \epsilon_\mathrm{\theta_0} \hfill \lor  k_\mathrm{\theta_\mathrm{0}}>k
    }
\end{equation}

\subsection{Analysis of Training Loss Landscape}
The training loss $E(\hat{\theta}_0)$ depends on the estimated circuit parameters $\hat{\theta}_0$, and the loss landscape provides a graphical representation which can be used to study the characteristics of optimization problems, such as estimation accuracy and convergence behavior \cite{liao2017theoryiilandscapeempirical}. The example loss curve is demonstrated using $\hat{L}$ and $\hat{R}_\mathrm{L}$ in {\color{black}Fig.~\ref{fig_method}(d), in which the remaining parameters in $\hat{\theta}_0$ are replaced with their actual values}. 

For the proposed method, the training loss landscape is analyzed primarily in terms of magnitude and smoothness. The magnitude of $E(\hat{\theta}_0)$ against $\hat{L}$ is at least one order of magnitude higher than that of $\hat{R}_\mathrm{L}$, causing  $\hat{L}$ to converge faster than $\hat{R}_\mathrm{L}$ during optimization, as the optimizer prioritizes parameters with the greatest loss reduction \cite{wright2006numerical}. 
Regarding smoothness, the $\hat{L}$ curve is smoother than the $\hat{R}_\mathrm{L}$ curve, resulting in lower variation for $\hat{L}$. 
{\color{black} For instance, the $E(\hat{\theta}_0)$ against $\hat{R}_\mathrm{L}$ curve exhibits higher oscillations between \SI{90}{\percent} to \SI{110}{\percent} of normalized $\hat{R}_\mathrm{L}$, creating local minima that can trap the optimizer and induce estimation errors\cite{li2018visualizing}. }


The above analysis is based on the converter model and the formulated optimization problem, corresponding to the second and third functional blocks in Fig.~\ref{fig_method}(a), and can therefore be applied to methods based on a similar problem formulation, for instance,  \cite{peng_digital_2021, zhao_parameter_2022, Gabriel2021LS, Sihui_3pInverter_2024}. 

There are also limitations of the loss landscape study. First, this analysis is case-dependent, as it relies on specific converter parameters and operating points, indicating the necessity of studying the training loss for each application.

Second, the two-dimensional loss landscape analysis only studies one parameter while the others remain fixed at their actual values, preventing a full examination of coupling effects between different parameters. 
For example, if $\hat{L}$ deviates from the actual $L$, the loss landscape in Fig.~\ref{fig_method}(d) may change. Therefore, the loss landscape study in this work is used solely to assess the difficulty of multiple parameter estimation and identify reliably estimated parameters for condition monitoring. More precise but computationally complex approaches can be explored through high-dimensional loss modeling \cite{B_ttcher_2024}.

\section{Simualtion Study}
The proposed method is validated through simulations under various load steps and noise levels. The circuit parameters and method settings are listed in Table~\ref{tab_circuit_param}. The sampling window covers the oscillation period of a load step, as the load step provides essential information for parameter identification~\cite{zhao_parameter_2022}. The sampling window length is \SI{1}{\ms} considering the estimation accuracy and computation burden.

\begin{table}[!t]\footnotesize 
\centering
\vspace{-8 pt}
\caption{Simulation Parameters.}\label{tab_circuit_param}
\begin{tabular}{cc|cc} 
\cline{1-4}\cline{1-4}
\textbf{Variable}  &    \textbf{Value}                       & \textbf{Variable} &\textbf{Value} \\ \cline{1-4}
Input volt.  $V_\mathrm i$     & \SI{48}{\V}      &   Inductance $L$                   &  \SI{1.40}{\mH}  \\ 
Output volt. $V_\mathrm o$ &    \SI{24}{\V} & Capacitance $C$          &   \SI{140}{\uF}  \\
Inductor ESR $R_\mathrm L$&    \SI{100}{\mohm} & Capacitor ESR $R_\mathrm c$&    \SI{300}{\mohm} \\
MOSFET  $R_\mathrm {dson}$&    \SI{40}{\mohm} & Diode  $V_\mathrm F$&    \SI{1}{\V} \\
Load resistance $R$              &    $\geq$\SI{2.88}{\ohm}  &    Max input power              &   \SI{200}{\W}  \\ 
Switching freq. $f_\mathrm{sw}$  & \SI{20}{\kHz}       &   Learning rate $L_\mathrm{Adam}$  & 0.025 \\ 
Sampling freq. $f_\mathrm{sa}$     &   \SI{40}{\kHz}   &    Learning rate $L_\mathrm{LBFGS}$ & 
0.05   \\
Prediction freq. $f_\mathrm{P}$     &   \SI{1}{\MHz}   &    Max epoch $k_\mathrm{Adam}$  &  250     \\
Sampling ratio $r_\mathrm{f}$  &   2 
& Max epoch $k_\mathrm{LBFGS}$  &  50 \\
 Range of $\hat{\theta}_0$  & $(0,\SI{500}{\percent}]$  & Tolerance $\epsilon_\mathrm{\theta_0}$  & 1e-6 \\ 
 Prediction size $M$  & $1000$ & Sampling size $N$  & $40$   \\ 
Current res. $i_\mathrm{Lres}$  & \SI{2.44}{\mA} & Voltage res. 
$v_\mathrm{ores}$  & \SI{7.32}{\mV}  \\ 
\cline{1-4}\cline{1-4}
\end{tabular}
\end{table}

\subsection{Estimation in Different Load Steps}

\begin{figure}[!t]
\centering
\includegraphics[width=3.2 in, page=3]{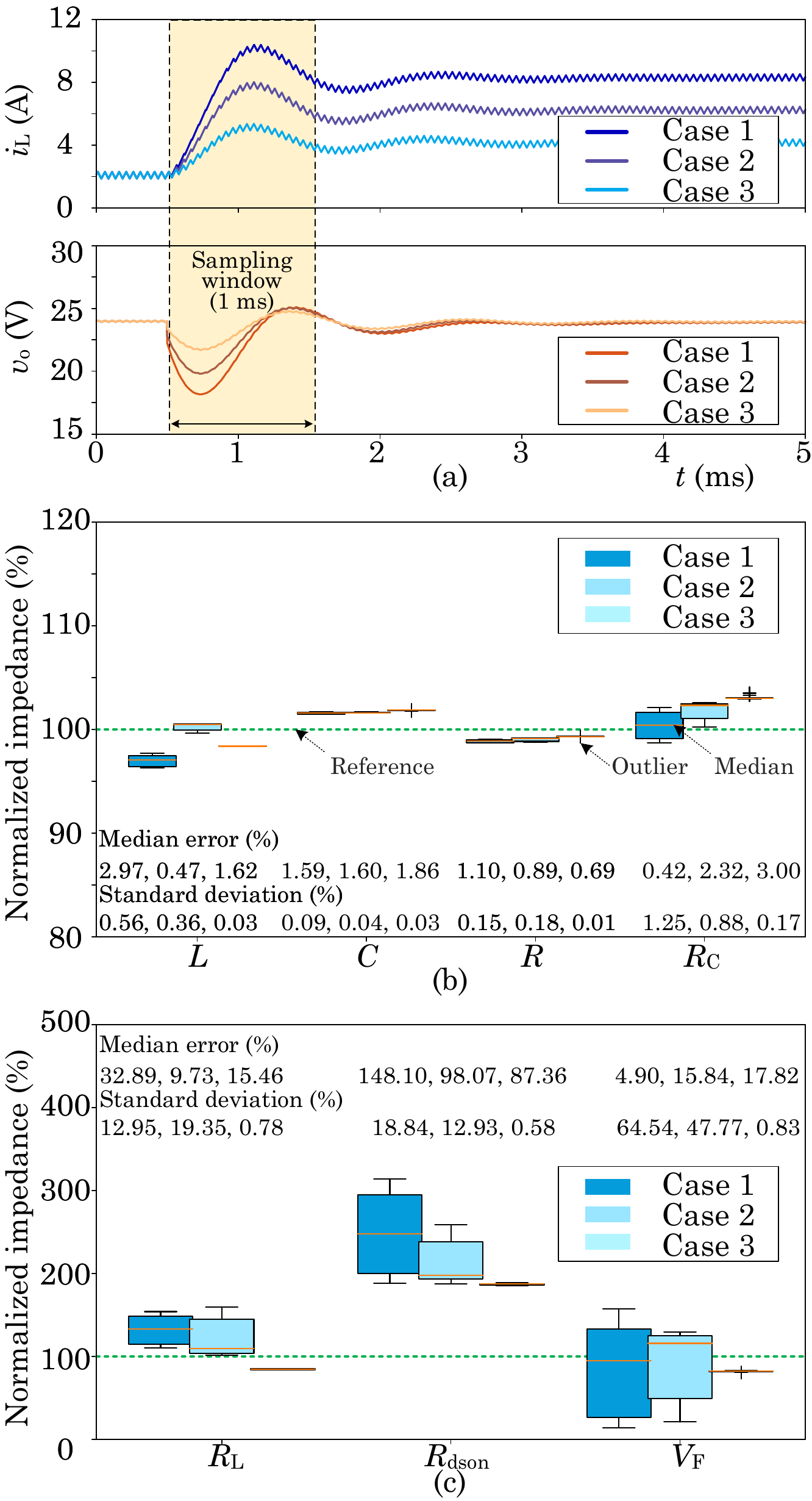}
\vspace{-8 pt}
\caption{Test Case 1 to 3: (a) Waveforms and estimations of three load steps. {\color{black} (b) Estimated circuit parameters $\hat{\theta}_1$.} {\color{black} (c) Estimated circuit parameters $\hat{\theta}_2$.}}
\label{fig_result_loadsteps}
\end{figure}

\begin{figure}[!t]
\centering
\includegraphics[width=3.2 in, page=6]{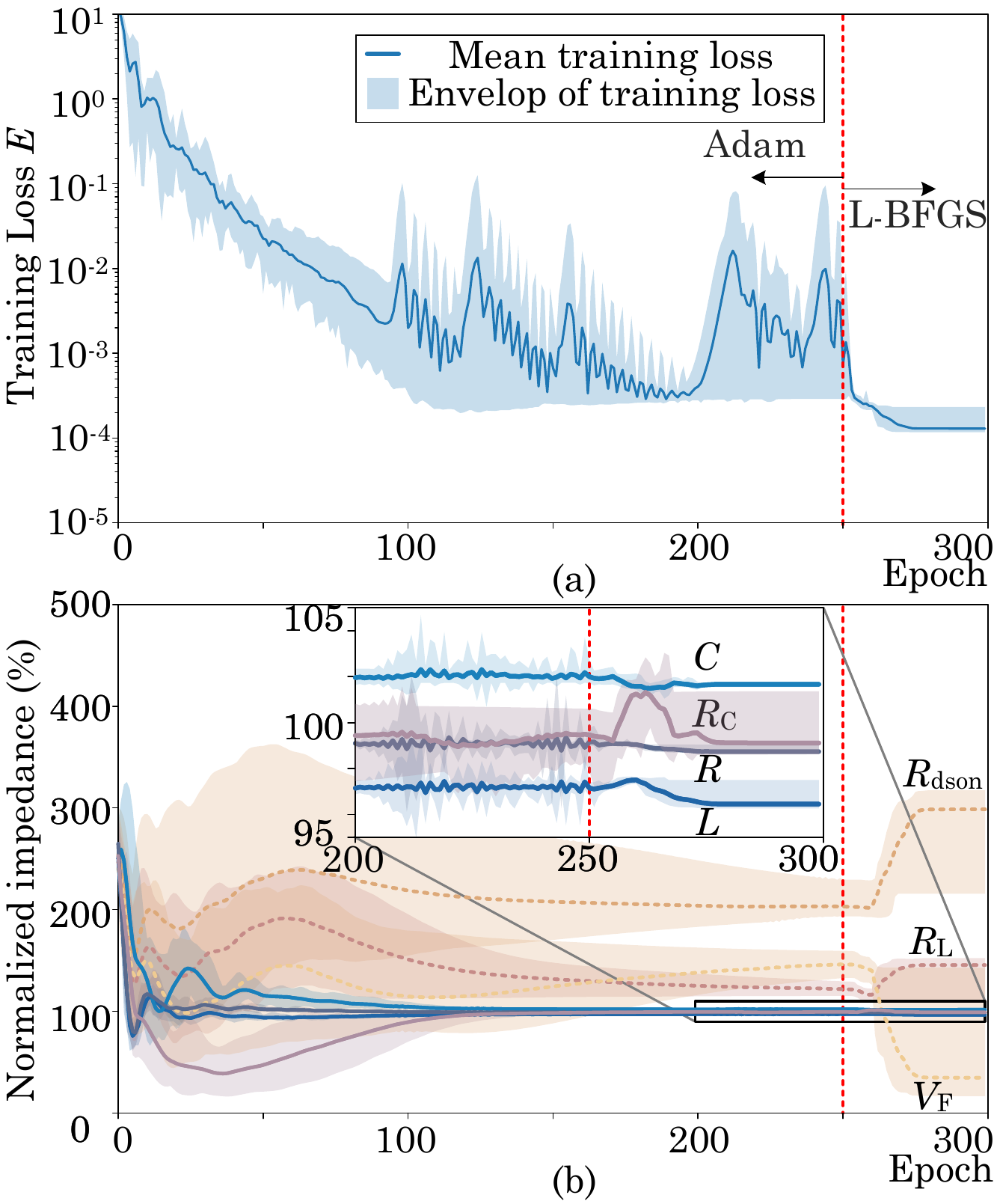}
\vspace{-8 pt}
\caption{As the epoch increases, (a) the training loss reduces gradually. (b) circuit parameters converge gradually.}
\label{fig_loss_loadsteps}
\end{figure}

The converter operates under varying load steps, making it essential to study their impact on estimation performance. Therefore, three load steps are designed, and the typical waveforms are illustrated in Fig.~\ref{fig_result_loadsteps}(a): 
\begin{itemize}
    \item \textbf{Case 1}: Load increases from \SI{25}{\percent} to \SI{100}{\percent}.
    \item \textbf{Case 2}: Load increases from \SI{25}{\percent} to \SI{75}{\percent}.
    \item \textbf{Case 3}: Load increases from \SI{25}{\percent} to \SI{50}{\percent}.
\end{itemize}
The estimated circuit parameters $\hat{\theta}_0$ across ten repetitive executions are normalized in Figs.~\ref{fig_result_loadsteps}(b) and (c), where $\hat{\theta}_1$ = $\{L$, $C$, $R$, $R_\mathrm{C}\}$ achieve both high accuracy (median estimation errors below \SI{3}{\percent}) and stable estimations (the standard deviations are below \SI{2}{\percent}). In contrast, $\hat{\theta}_2$ = $\{R_\mathrm{L}$, $R_\mathrm{dson}$, $V_\mathrm{F}\}$ exhibit lower accuracy (with a maximum error of \SI{148}{\percent}) and greater variation (with a maximum standard deviation of \SI{65}{\percent}). {\color{black}These results indicate that $\hat{\theta}_1$ performs consistently across tested load steps, whereas $\hat{\theta}_2$ fails to converge in three cases.}

The convergence trend of Case 1 is demonstrated in Figs.~\ref{fig_loss_loadsteps}(a) and (b). The training loss $E(\hat{\theta}_0)$ gradually decreases as the number of training epochs increases. 
The rapid decline in $E(\hat{\theta}_0)$ during the initial phase (epoch $<$ 100) is mainly due to the convergence of $\hat{\theta}_1$ toward the actual parameter, \SI{100}{\percent} normalized impedance. Subsequently, $\hat{\theta}_2$ begins to converge after switching from the Adam optimizer to L-BFGS at epoch 250, as highlighted by the red dashed line. L-BFGS also stabilizes the estimation of $\hat{\theta}_1$, as shown in the zoomed-in figure.


\begin{figure}[!t]
\centering
\includegraphics[width=3.2 in, page=4]{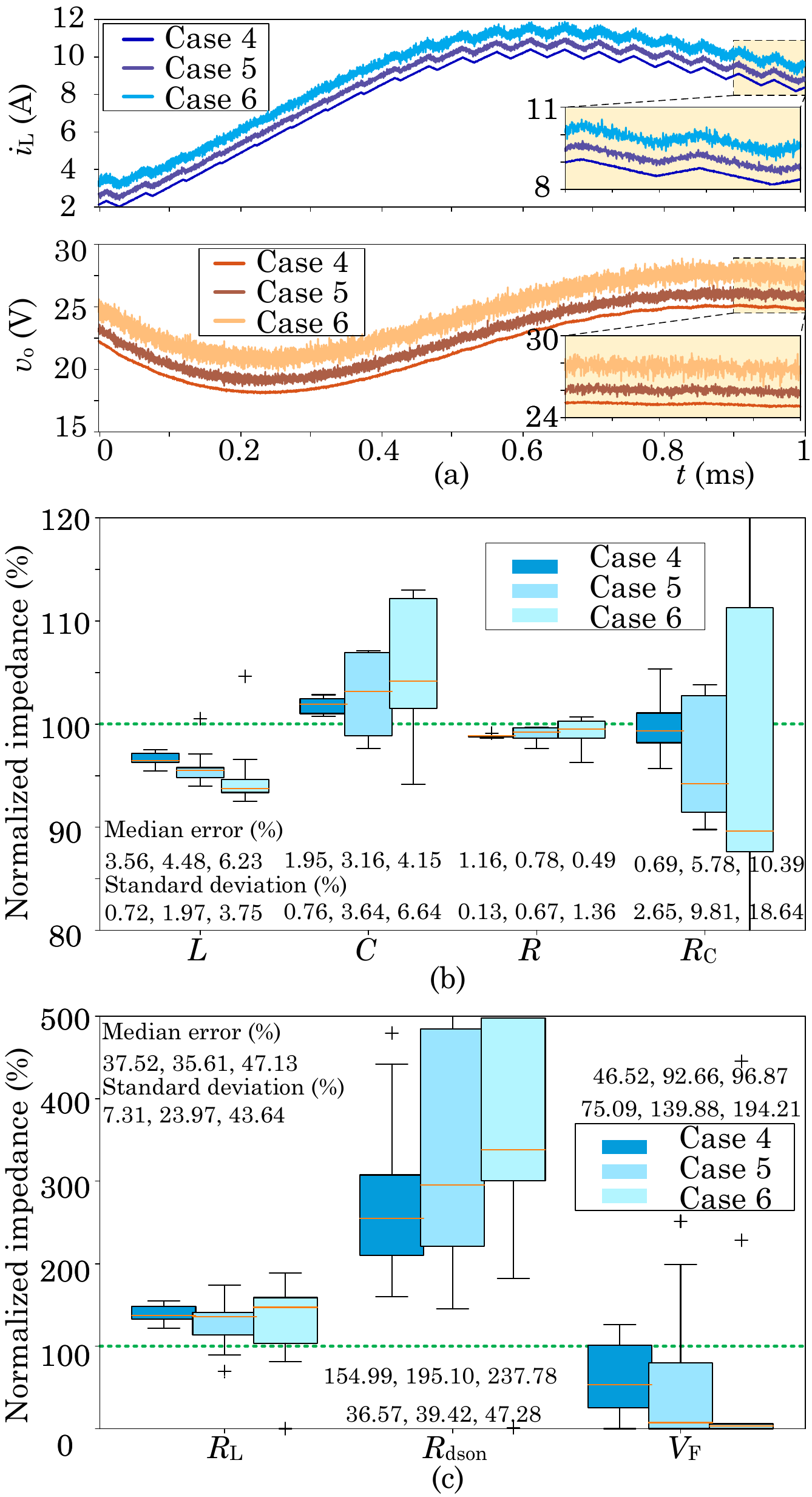}
\vspace{-8 pt}
\caption{Test Case 4 to 6: (a) The waveforms of state vectors with different noise levels, where a small offset is added to each signal to separate signals. (b) Boxplots of estimated $\hat{\theta}_1$. (c) Boxplots of estimated $\hat{\theta}_2$.}
\label{fig_result_noises}
\end{figure}

\begin{figure}[!t]
\centering
\includegraphics[width=3.2 in, page=5]{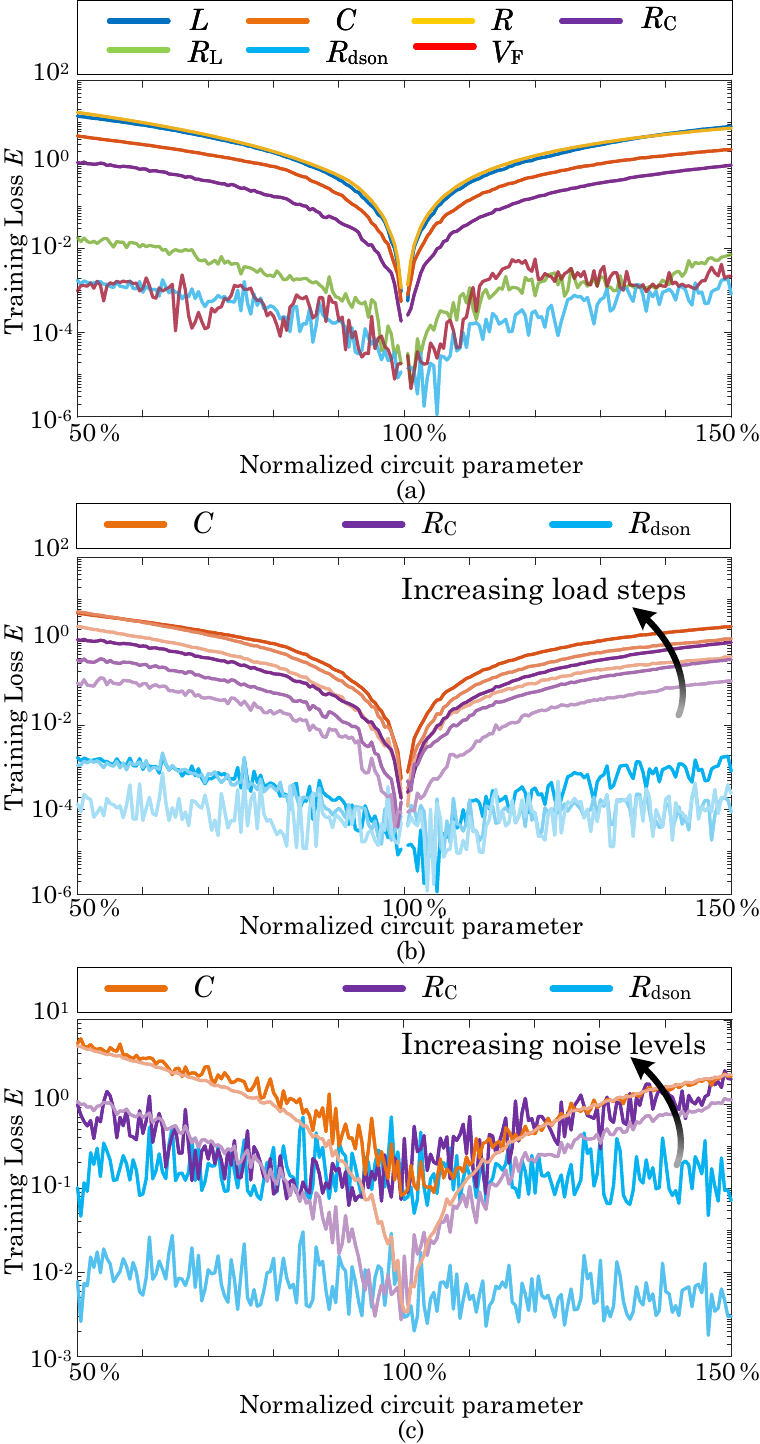}
\vspace{-8 pt}
\caption{Training loss landscape of- (a) all circuit parameters in Case 1; (b) selected circuit parameters across different load step sizes, from Case 1 to 3; (c) selected circuit parameters across different noise levels, from Case 4 to 5. }
\label{fig_loss1}
\end{figure}


\subsection{Estimation with Noise}

This section analyzes noise immunity by introducing different levels of noise into the measured state vector. The noise follows a normal distribution with zero mean and a standard deviation proportional to the analog-to-digital conversion (ADC) resolution \cite{zhao_parameter_2022}. {\color{black}The noise levels are case dependent and defined as ten in \cite{zhao_parameter_2022}. To study the extreme condition, their range is selected as from five to fifty times of the resolution: }
\begin{itemize}
    \item \textbf{Case 4}: The noise level is 5 times the resolution.
    \item \textbf{Case 5}: The noise level is 25 times the resolution.
    \item \textbf{Case 6}: The noise level is 50 times the resolution.
\end{itemize}
where the resolutions of the inductor current $i_\mathrm{Lres}$ and output voltage $v_\mathrm{ores}$ are calculated using a 12-bit ADC with measurement ranges of \SI{10}{\A} and \SI{30}{\V}, respectively. The waveforms of the state vector with noises are illustrated in Fig.~\ref{fig_result_noises}(a), with a small offset added for clearer visualization.

The estimations are illustrated in a boxplot in Figs.~\ref{fig_result_noises}(b) and (c), showing that the errors of $\hat{\theta}_1$ remain below \SI{11}{\percent}. Noise immunity varies across different parameters, where the standard deviations of $L$, $C$, $R$ remain consistently below \SI{5}{\percent}; while the standard deviation of $R_\mathrm{C}$ increases significantly from \SI{2.65}{\percent} to \SI{18.64}{\percent} as the noise rises from Case 4 to Case 6. The estimations of $\hat{\theta}_2$ are also noise-dependent, making them difficult to estimate accurately and stably. 
The reasons for these variations in noise immunity are explained in the following section.




\subsection{Loss Landscape Analysis}
This section analyzes the training loss landscape to understand circuit parameter estimations from two aspects:
\begin{itemize}
    \item The different challenges in estimating $\hat{\theta}_1$ and $\hat{\theta}_2$.
    \item The influence of operating points and load steps on method performance.
\end{itemize}

As illustrated in Fig.~\ref{fig_loss1}(a), the training loss $E(\hat{\theta}_0)$ depends on the circuit parameters. The test condition is Case 1 with no noise. 

The loss landscape is explained from three aspects: magnitude, smoothness, and optimal solution. 
\begin{itemize}
    \item \textbf{Magnitude}: The training loss curve of $\hat{\theta}_1$ is at least one order of magnitude higher than $\hat{\theta}_2$, causing $\hat{\theta}_1$ to converge faster, as observed in the convergency trend of  Fig.~\ref{fig_loss_loadsteps}(b).
    \item \textbf{Smoothness}: The loss curves of $\hat{\theta}_1$ are smoother than that of $\hat{\theta}_2$, resulting in lower variation for $\hat{\theta}_1$, as shown in Fig.~\ref{fig_result_loadsteps}(b).
    \item \textbf{Optimal Solution}: The optimal solution for most circuit parameters aligns with their actual values (\SI{100}{\percent}), which differs from estimations of $\hat{\theta}_2$ in Fig.~\ref{fig_result_loadsteps}(c). The discrepancy is mainly due to the limitation of neglecting the coupling effects between multiple circuit parameters. 
\end{itemize}


The training loss also varies with operating conditions, as illustrated in Fig.~\ref{fig_loss1}(b). {\color{black} Three} circuit parameters $C$ and $R_\mathrm{C}$, and $R_\mathrm{dson}$ are analyzed, representing a solvable parameter, an operating-condition-dependent parameter, and a nonsolvable parameter, respectively. 

As the load step increases, the training loss curve of $C$ remains consistent, while the magnitude of $R_\mathrm{C}$ increases for one order of magnitude. 
For $R_\mathrm{dson}$, the loss curve remains lower than the others and oscillatory, preventing the estimations from converging regardless of the load step used. 

The noise level further complicates the estimation process. Higher noise introduces oscillations and flattens the loss curves around the \SI{100}{\percent} point, leading to increased error and variation in $C$, $R_\mathrm{C}$ and $R_\mathrm{dson}$ in Figs.~\ref{fig_result_noises}(b) and (c).

In summary, the loss landscape analysis reveals distinct performance characteristics among three types of parameters. 
\begin{itemize}
    \item \textbf{Accurately Estimated Parameters} (e.g., $L$, $C$, $R$): These parameters exhibit high accuracy and low variation across different operating conditions, making them suitable for condition monitoring \cite{Soliman_2016_CM_review}.
    \item \textbf{Condition-Dependent Parameters} (e.g., $R_\mathrm{C}$): While estimable, their accuracy and variation are strongly influenced by operating conditions and noise levels. Using this term as a health indicator requires additional caution to mitigate operational and noise effects.
    \item \textbf{{\color{black}Unreliable Parameters}} (e.g., $R_\mathrm{L}$, $R_\mathrm{dson}$, $V_\mathrm{F}$): These parameters cannot be reliably estimated under the given problem formulation, making them unsuitable for condition monitoring. 
\end{itemize}


\subsection{Method Comparison}
{\color{black}The proposed method is compared with another PINN \cite{zhao_parameter_2022} in terms of method and performances in Table~\ref{tab_method_comparison}.

Both PINNs share a similar structure, consisting of a physics model, numerical solver, and optimizer. The physics models are identical, using the estimated circuit parameters to build a state-space model. Whereas the numerical solvers differ: the method in \cite{zhao_parameter_2022} employs an implicit Runge-Kutta solver, while the proposed method uses the Forward Euler method. Regarding the optimizer, both methods use the Adam and L-BFGS optimizers to update the weights and biases of the neural network. The key difference lies in the network size--the structure in \cite{zhao_parameter_2022} is larger than the proposed method. This difference is due to the output layer in \cite{zhao_parameter_2022},
which determines the number of hidden states in a switching cycle and is set to be at least 20 to minimize truncation error. 
In contrast, the proposed neural network decouples its outputs from the hidden states, resulting in a smaller and more efficient structure. The difference in size is further reflected in the model size, which are \SI{49}{kB} and \SI{2}{kB}, respectively. 

These two methods are compared in Case 1 on an Intel Core i7-1165G7
@2.8 GHz, 32-GB RAM, 64-bit system. Each method is executed ten times, and the results are averaged. 

The maximum epoch and convergence time of the proposed method are significantly smaller than those of \cite{zhao_parameter_2022}, due to its smaller network structure. However, the estimations of $\theta_1$ exhibit a slightly lower accuracy and higher variations. Therefore, the proposed method is better suited for scenarios with limited memory and computational resources. The proposed method offers sufficient accuracy for inductor and capacitor condition monitoring, with end-of-life criteria being \SI{20}{\percent}\cite{Zhaoyang2021_C_CM_review}. If accuracy and low-variation are the primary concerns, the method in \cite{zhao_parameter_2022} is more suitable while requiring more computational resources.
}

\begin{table}[!t]\footnotesize 
\label{tab_method_comparison}
\caption{Comparing the proposed and existing PINN methods.}\label{tab_method_comparison}
\begin{tabular}{ccc}
\hline
\cline{1-3}\cline{1-3}
\multicolumn{1}{c}{Method}                        & \multicolumn{1}{c}{PINN \cite{zhao_parameter_2022}}                               & \textbf{Proposed method}                   \\ \hline
\multicolumn{3}{c}{Method comparison}                              \\ \hline
\multicolumn{1}{c}{Physics model}                 & \multicolumn{1}{c}{State-space model}       & \multicolumn{1}{c}{State-space model}                                     \\ 
\multicolumn{1}{c}{Solver} & \multicolumn{1}{c}{IRK}                             & FE                   \\ 
\multicolumn{1}{c}{Optimizer}                     & \multicolumn{1}{c}{NN+Adam+LBFGS}&    \multicolumn{1}{c}{NN+Adam+LBFGS}                                  \\ 
\multicolumn{1}{c}{NN structure}             & \multicolumn{1}{c}{[5$\times$50$\times$50$\times$50$\times$50$\times$50$\times$40]} & [2$\times$16$\times$16$\times$7]    \\ 
\multicolumn{1}{c}{NN output}                & \multicolumn{1}{c}{Hidden state vector} & Estimated parameters \\ 
\multicolumn{1}{c}{Network size}                & \multicolumn{1}{c}{\SI{49}{kB}}       & \SI{2}{kB} \\ \hline
\multicolumn{3}{c}{Result comparison}                                                                                           \\ \hline
\multicolumn{1}{c}{Max epochs}                    & \multicolumn{1}{c}{250000}                          & 300                  \\ 
\multicolumn{1}{c}{Converge time}              & \multicolumn{1}{c}{\SI{243}{\s}}                           & \SI{130}{\s}                 \\ 
\multicolumn{1}{c}{Error of $\hat{\theta}_1$}        & \multicolumn{1}{c}{ \SI{0.11}{\percent}}                                &         \SI{1.84}{\percent}             \\ 
\multicolumn{1}{c}{Variation of $\hat{\theta}_1$}       & \multicolumn{1}{c}{ \SI{0.04}{\percent}}                                &         \SI{1.06}{\percent}             \\ 
\cline{1-3}\cline{1-3}
\end{tabular}
\begin{tablenotes}
    \item[*] *PINN: Physics-informed neural network; IRK: Implicit Runge Kutta; FE: Forward Euler; NN: Neural network; Adam: adaptive moment estimation; Limited-memory Broyden–Fletcher–Goldfarb–Shanno. 
\end{tablenotes}
\end{table}



\section{Conclusion}
This work proposes a PINN with three functional blocks: parameter estimation, state vector prediction, and loss calculation. The proposed method effectively identifies the circuit parameters of the buck converter, with median estimation errors below \SI{7}{\percent}, \SI{5}{\percent}, \SI{2}{\percent}, \SI{11}{\percent} for $L$, $C$, $R$, $R_\mathrm{C}$ in test Cases 1 to 6, respectively. Compared with an existing PINN method, the proposed method offers the advantage of a smaller neural network, lower memory requirements, and faster convergence. 

As for the other circuit parameters that cannot be estimated accurately, the training loss landscape is analyzed to evaluate the difficulty in multiple parameter estimation with different load steps and noise levels. The buck converter parameters are then classified into three categories according to their feasibility of being a health indicator, which can be used to guide the implementation of system-level condition monitoring.

\section*{Acknowledgments}
This project is supported by the European Union’s Horizon 2020 research and innovation program under the Marie Skłodowska-Curie grant agreement No. 955614. 


\bibliographystyle{IEEEtran}
\bibliography{C4ref}

\vfill
\end{document}